\documentclass[a4paper]{article}
\usepackage[dvips]{graphicx}

\usepackage{amssymb}
\newtheorem{law}{Proposition}

\begin{document}

\title{Wavelet Quantum Search Algorithm with Partial Information}

\author{Sangwoong Park{\footnote{Email address: sangwoong@ihanyang.ac.kr}}
,Joonwoo Bae{\footnote{Email address: jwbae@newton.hanyang.ac.kr}}, Younghun Kwon{\footnote{ Email address: yyhkwon@hanyang.ac.kr}} \\Departments of Mathematics$^{*}$ and Physics$^{\dag\ddag}$\\ Hanyang University \\Kyunggi-do, Ansan, South Korea}

\maketitle
\vspace{-.4in}
\begin{abstract}
It is questionable that Grover algorithm may be more valuable than a classical one, when a partial information is given in a unstructured database. In this letter,  to consider quantum search when a partial information is given, we replace the Fourier transform in the Grover algorithm with the Haar wavelet transform. We then, given a partial information $L$ to a unstructured database of size $N$, show that there is the improved speedup, $O(\sqrt{N/L})$. 
\end{abstract}
 
\bigskip
\bigskip

Suppose that we have a problem of finding a desired one of unstructured $N$ items. It is known that a classical search algrithm may take $O(N)$ times, but the fast quantum search algorithm provides the quadratic speedup, $O(\sqrt{N})$.\cite{2} The quantum algorithm whose central idea is amplitude amplification was first provided by Grover in 1996.\cite{1}\cite{3} It is well-known that Grover algorithm is 1)optimal in the context of applying unitary operators repeatdly and 2)efficient in searching a target of a unstructured database. However, it is not evident that, when a partial information is given, the Grover algorithm is still more valuable than a classical one. In this letter, we provide the quantum search algorithm which is able to benefit from a partial information. The key building block in our construction is the (Haar) wavelet transform instead of the Fourier transform in the Grover algorithm. \\
 Let us first consider the Grover algorithm. A bijection between a database and quantum states is necessary before applying the Grover algorithm. If a superposition of $N$ states is initially prepared, the Grover algorithm amplifies the amplitude of the target state up to around one, while those of other states dwindle down to nearly zeros. The amplitude amplification is performed by two inversion operations: inversion about the target by the oracle and inversion about the initial state by the Fourier transform. Noting the fact that two simultaneous reflections about two mirrors crossing by an angle $\alpha$ induce $2\alpha$ rotation, one may imagine that the inversions in the Grover algorithm rotate the initial state around the target state.\cite{3}
If the target state and the initial state are denoted by $|w\rangle$ and  $|\psi\rangle$ respectively,(here the initial state is prepared by the Fourier transform of a state $|k\rangle$, i.e. $|\psi\rangle = \mathcal{F}|k\rangle$) the inversion operators is expressed as

\begin{eqnarray}
O_{|w\rangle} & = & 1-2|w\rangle\langle w| \nonumber \\
I_{|\psi\rangle} & = & 1-2|\psi\rangle\langle \psi| \nonumber \\
\end{eqnarray}

Since $I_{|\psi\rangle} = \mathcal{F}I_{|k\rangle}\mathcal{F}^{\dagger}$, the Grover operator is written as

\begin{eqnarray}
U = -\mathcal{F}I_{|k\rangle}\mathcal{F}^{\dagger}O_{|w\rangle} \nonumber \\
\end{eqnarray}

 Then, after applying the operator $O(\sqrt{N})$ times, the final state comes to 

\begin{eqnarray}
|final \rangle = U^{ O(\sqrt{N})} \mathcal{F}|k\rangle \nonumber\\
\end{eqnarray}

The probability to obtain the target state is $Prob(w) = |\langle w |final \rangle|^{2}$, which is $1-\epsilon^{2}$, $\epsilon \ll 1$. The query complexity of this algorithm, the number of callings of the oracle, is therefore $O(\sqrt{N})$. We here note that the running time has nothing to do with the choice of $|k\rangle$.\\

To consider the partial information,   let us think about the following situation.\\

{\it Suppose that Hanyang university library has $N$($=2^{n}$, for some $n \in \mathbb{Z}$) books. Each
book has a code number identifying itself. Recently, the library decides to label new code numbers to each book since a better 
way to classify books has provided. A labelling machine is applied to the tedious job. Each code is composed of $n$ numbers. The first numbers must
be $1$, since they stands for the region. And Hanyang university is
in South Korea, so the second number is $1$.The other
numbers stand for the category each book is included. An earlier
number means a larger category. The rules of labeling a code
number is that the $\lambda$-th number, $j$, is in
$[1,2^{\lambda-1}]$ for $\lambda \geq 1$. The following codes are
a good example,

\begin{eqnarray}
&&11247143040123 \nonumber\\
&&11111101164001 \nonumber \\
\end{eqnarray}

The first code means $1,1,2,4,7,14,30,40,123$ and the next one
$1,1,1,1,1,10,11,64,001$. The first two numbers $1,1$ stands for
the region, and Hanyang University. To aid the searching a book in the library, the library also has implemented the Grover algorithm in a quantum computer. \\
After the labelling is completed, one discovers a
fault in the machine. Therefore, the library decides to do labelling again, fixing the fault in the labelling machine. They then recommend to apply the Grover algorithm to whom may be concerned to search a book, during labelling. \\
Dr. Lee, a postdoc in physics, has to submit his research paper by tomorrow $12$ a.m. It is now 6 p.m. He wish to fill the reference section in the paper, but the library in mess blocks his work. The director of the library recommends him to try the Grover algorithm, but it was calculated ,based on the number of books $N$, that $24$ hours is the running time of the algorithm. Therefore, he concludes that the Grover algorithm cannot help him. At that moment, he gets a call from the library that the only $\lambda$-th number of all code numbers is correct.}\\

It is sure that the Grover algorithm cannot complete the search task in time, since it takes over $24$ hours. The only thing that Dr. Lee can use in order to overcome his hopeless situation, is the fact that only the $\lambda$-th number  was correctly labelled. The partial information may save his problem. However, the Grover algorithm cannot benefit from the partial information of this problem.\\
We here introduce the fast wavelet quantum search algorithm(WQSA), which is a modification of the Grover algorithm by replacing the Fourier transform with the Haar wavelet transform, to resolve the situation of Dr. Lee. Let us note that to apply the following operator

\begin{eqnarray}
U_{\mathcal{W}} = -\mathcal{W}^{\dagger}I_{|k\rangle}\mathcal{W}O_{|w\rangle} \nonumber \\
\end{eqnarray}

is one iteration of the WQSA. The Haar wavelet transform $\mathcal{W}$ is represented $\mathcal{W}= \mathcal{W}_n \mathcal{W}_{n-1} \cdots \mathcal{W}_1$ where

\begin{eqnarray}
\mathcal{W}_{k} =  \left[
\begin{array}{ccc}
H_{2^{n-k+1}}& O_{2^{n-k+1} \times (2^n-2^{n-k+1})} \\
O_{(2^n-2^{n-k+1}) \times 2^{n-k+1}} & I_{2^n-2^{n-k+1}}
\end{array} \right].
\end{eqnarray}

and $ H_{2^k} $ the Haar 1-level decomposition operator as follows;


\begin{eqnarray}
H_{2^k} = \frac{1}{\sqrt{2}} \left[
\begin{array}{cccccccccccc}
       1 & 1 &  0 &  0&\dots & 0      \\
       0 &  0 & 1 & 1 & 0 & \vdots      \\
       \vdots & \vdots  & & \ddots&0 & 0  \\
       0 & 0 &\dots& 0 & 1 & 1          \\
       1 & -1 &  0 &  0&\dots & 0       \\
       0 &  0 & 1 & -1 & 0 & \vdots     \\
       \vdots & \vdots  & & \ddots&0 & 0 \\
       0 & 0 &\dots& 0 & 1 & -1
\end{array} \right]_{2^k \times 2^k}
\end{eqnarray}

We have used $I_n$ as the $n \times n$ unit matrix and $O_{n,m}$ as the $n \times
m$ zero matrix. It is clear that the wavelet transform $\mathcal{W}$ is unitary since the operator $H_{2^{k}}$ is unitary. \\
Since the operator is composed of the wavelet transform, consistently, the initial state is prepared by applying the inverse wavelet transform $\mathcal{W}^{\dagger}$ to a state $|k\rangle$, i.e., the initial state is now $|\psi\rangle = \mathcal{W}^{\dagger}|k\rangle$. The power of our WQSA appears in the initialization procedure. It is quite remarkable that the state $\mathcal{W}^{\dagger}|k\rangle$ is a superposition of $N/L$ states, where $L=2^{\lambda-1}$($\lambda$ is given by $k$), while the state $\mathcal{F}|k\rangle$ is a superposition of $N$ states.  Then it is expected that the running time is $O(\sqrt{N/L})$. Choosing the initial state as $\mathcal{W}^{\dagger}|k\rangle$, $k\neq 0,1$ when the target state exists in the restricted domain of the $N/L$ states, we look forward to an improved speedup with the partial information. Since $k\in\lbrace 3,4,\cdots,N(=2^{n})-1\rbrace$, by setting $k=2^{\lambda-1}+j$, $1\leq j \leq 2^{\lambda}$ and $\lambda \geq 1$, and $N_{1} = 2^{n-\lambda +1}(=N/L)$, the state $\mathcal{W}^{\dagger}|k\rangle$ is explicitly,

\begin{eqnarray}
\mathcal{W}^{\dagger}|k\rangle = \sum_{\alpha =(j-1)N_{1}}^{(j-1)N_{1}+N_{1}/2-1}|\alpha\rangle -\sum_{\beta=(j-1)N_{1}+N_{1}/2}^{jN_{1}-1}|\beta\rangle \nonumber \\
\end{eqnarray}

The following diagram shows the set of states composing $\mathcal{W}^{\dagger}|k\rangle$.

\begin{figure}[here]
\begin{center}
\includegraphics[width=0.4\textwidth]{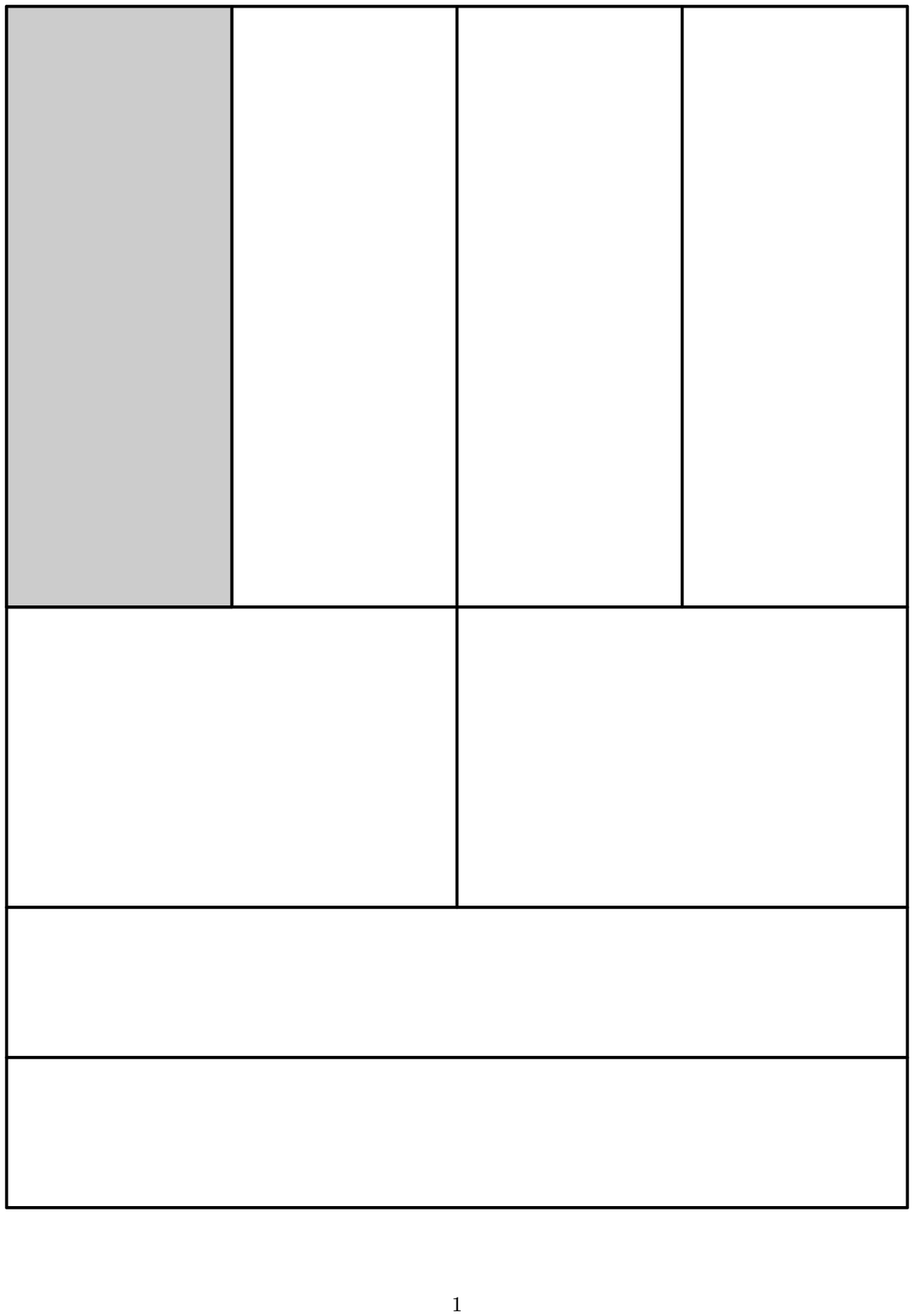}
\end{center}
\caption{For N=8, each rectangular represents a set of states composing $\mathcal{W}^{\dagger}|i\rangle$. The $x-$axis runs from $0$ to $8$. When the initial state is $\mathcal{W}^{\dagger}|0\rangle$ or $\mathcal{W}^{\dagger}|1\rangle$, the running time is the same to that of the Grover algorithm since the lowest two retangulars include all states.}
\end{figure}

We thus arrive at the following proposition.

\begin{law}[Wavelet Quantum Search]
Suppose that we solve the problem of finding a desired one in the set $\mathcal{A} = \lbrace |a\rangle ~|~a = 0,1,2,3,\cdots, 2^{n-1}~\rbrace$. Given a partial information that the target state is in the subset $\mathcal{A}_{\lambda}^{j} = \lbrace |z\rangle ~|~ (j-1)2^{n-\lambda}\leq z \leq j2^{n-\lambda}-1, ~~1\leq j\leq 2^{\lambda} \rbrace$, we complete the search task in $O(\sqrt{2^{n-\lambda+1}})$ times by choosing the initial state as $\mathcal{W}^{\dagger}| 2^{\lambda -1}+j\rangle$.
\end{law}

proof) Let the target state $|w\rangle \in \mathcal{A}_{\lambda}^{j}$. Let us take the initial state as $\mathcal{W}^{\dagger}|2^{\lambda-1}+j \rangle$. It suffices to show that it takes $O(\sqrt{2^{n-\lambda+1}})$ times for the WQSA to find the target state with the following setting.\\
Let $N_{1} = 2^{n-\lambda+1}$. The wavelet quantum search operator is

\begin{eqnarray}
U_{\mathcal{W}} = -\mathcal{W}^{\dagger}I_{|k\rangle}\mathcal{W}O_{|w\rangle} \nonumber \\
\end{eqnarray}

where $\mathcal{W}$ is the Haar wavelet transform. Applying the
operator $\mathcal{W}^{\dagger}$ to the $|k\rangle$, we have the initial state

\begin{eqnarray}
|\psi\rangle = \sum_{\alpha =(j-1)N_{1}}^{(j-1)N_{1}+N_{1}/2-1}|\alpha\rangle -\sum_{\beta=(j-1)N_{1}+N_{1}/2}^{jN_{1}-1}|\beta\rangle \nonumber \\
\end{eqnarray}

, which can be rewritten as follows

\begin{eqnarray}
|\psi\rangle &=& \frac{\epsilon_{w}}{\sqrt{N_{1}}}|w\rangle  + \epsilon_{r} \sqrt{\frac{N_{1}-1}{N_{1}}}|r\rangle \nonumber\\ 
\end{eqnarray}

where $\epsilon_{i}\in\lbrace \pm 1 \rbrace$ and the state 

\begin{eqnarray}
|r\rangle &=& \sqrt{\frac{1}{N_{1}-1}}\sum_{\gamma \neq w}\epsilon_{\gamma}|\gamma\rangle \nonumber \\
\end{eqnarray}

 is the orthogonal complement of the target state. The $m$ iterations of the operator $U_{\mathcal{W}}$ create the following state,

\begin{eqnarray}
|\psi_{m}\rangle &=& U^{m}|\psi\rangle \nonumber \\
\end{eqnarray}

The probability to obtain the target state after the $m$ iterations is

\begin{eqnarray}
P_{m} = |\langle w|\psi_{m}\rangle |^{2}&=& \cos^{2}(m\theta - \varphi) \nonumber\\
\end{eqnarray}

,where $\theta = \sin^{-1}(2\epsilon_{w}\epsilon_{r}\sqrt{N_{1}-1}/N_{1})$ and $\varphi = \cos^{-1}(\epsilon_{w}/N_{1})$. Hence, we have shown that the total number of iterations is $O(\sqrt{2^{n-\lambda+1}})$. If we denote $N=2^{n}$ and $L=2^{\lambda-1}$, then the running time is written as $O(\sqrt{N/L})$ ~~~~~Q.E.D.\\

Let us revisit the Dr. Lee's problem.The partial information that the $\lambda$-th number $j$ is correctly labelled leads Dr. Lee to apply the WQSA so that the reference section is filled in time. However note that there is no improvement in running time when the intial state is $\mathcal{W}^{\dagger}|0\rangle$ or $\mathcal{W}^{\dagger}|1\rangle$ since, in those cases, the initial state is still a superposition of $N$ states. Therfore, from the proposition, we know that he can complete the submission in time if the $\lambda$ is larger than $2$. \\
We have exerted how to utilize a partial information, in order to enhance quantum search. Our construction provides a way for quantum search to benefit from a partial information. Since the running time of the Grover algorithm has nothing to do with the choice of unitary operator, the complexity of the WQSA is the same to the Grover algorithm. However, we have obtained the speedup $O(\sqrt{N/L})$ by preparing the initial state as $\mathcal{W}^{\dagger}|k\rangle$. The running time of the WQSA depends on the choice of $k$, while that of the Grover algorithm does not. This is because the state $\mathcal{W}^{\dagger}|k\rangle$ is a superpositin of states in the restricted domain of $N/L$ states. The speedup is indeed originated in the initialization. \\
 Finally, let us discuss the Haar wavelet basis. Although other wavelet transforms may be applied to the WQSA, we chose the Haar wavelet transform. 

\begin{figure}[here]
\begin{center}
\includegraphics[width=1\textwidth]{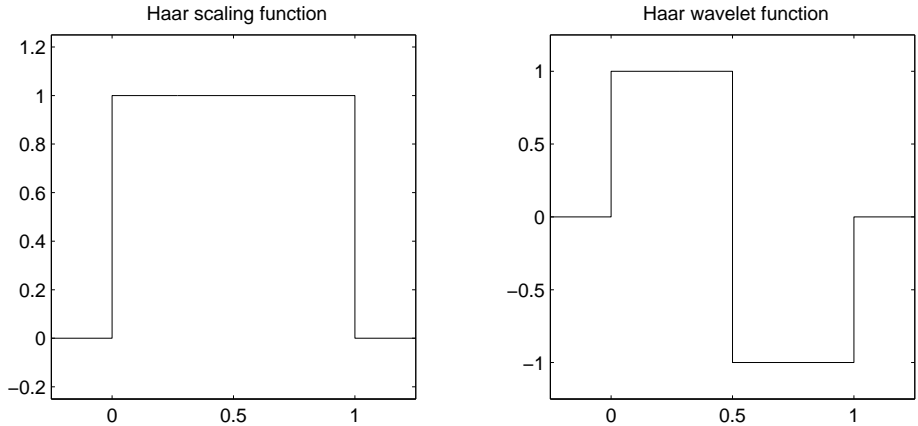}
\end{center}
\caption{The father wavelet(left) and the mother wavelet(right) of the Haar basis.}
\end{figure}
\vspace{0.5cm}

It is observed that the first half of the Haar wavelet basis differs with the second half of the wavelet basis on the phase $e^{i\pi}$. This implies that destructive and constructive interference between states accepts a set of states containing the target and rejects the other states. In this sense, other known wavelet basis, e.g. Daubechies's, are not appropriate to play the role of seclecting a subset of the $N$ states.

\section*{Acknowledgement}
J. Bae is supported in part by the Hanyang University Fellowship and Y. Kwon is supported in part by the Fund of Hanyang University.


\begin{thebibliography}{99}


\bibitem{1} G. Brassard, P. Hoyer, M. Mosca and Alain Tapp , Quantum Amplitude Amplification and Estimation, LANL Report No. quant-ph/0005055

\bibitem{2} L. K. Grover, Phys. Rev. Lett. $\mathbf{79}$, 325 (1997)

\bibitem{3} R. Jozsa, Searching in Grover's Algorithm, LANL Report No. quant-ph/9901021

\end{thebibliography}
\end{document}